\documentclass[10pt, conference, letterpaper, bottom=0.5in, top=0.7in]{IEEEtran}

\IEEEoverridecommandlockouts
\usepackage{lipsum}
\usepackage{hyperref}
\usepackage{scalerel}
\usepackage{tikz}
\usetikzlibrary{svg.path}
\usepackage{xcolor}
\usepackage{cite}
\usepackage{algorithm}
\usepackage{algpseudocode}
\usepackage{stackengine}
\usepackage{amsmath,amssymb,amsfonts}
\usepackage{graphicx}
\usepackage{textcomp}
\usepackage{xcolor}
\usepackage[utf8]{inputenc}
\usepackage{fixmath}
\usepackage{balance}
\usepackage{multirow}
\usepackage{xparse}
\NewDocumentCommand{\Log}{o}{%
  \IfNoValueTF{#1}{}{{}^{#1}\!}\log}%
\usepackage{array}
\usepackage{booktabs}
\usepackage{soul}
\usepackage{lineno}
\usepackage{inputenc}
\usepackage{subfigure}
\usepackage{subcaption}
\usepackage{authblk}
\usepackage[export]{adjustbox}
\usepackage{mathtools}
\usepackage{cuted}
\usepackage{gensymb}
\usepackage{url}
\usepackage{breakurl}

\def\BibTeX{{\rm B\kern-.05em{\sc i\kern-.025em b}\kern-.08em
    T\kern-.1667em\lower.7ex\hbox{E}\kern-.125emX}}

\definecolor{lime}{HTML}{A6CE39}
\DeclareRobustCommand{\orcidicon}{%
	\begin{tikzpicture}
	\draw[lime, fill=lime] (0,0) 
	circle [radius=0.16] 
	node[white] {{\fontfamily{qag}\selectfont \tiny ID}};
	\draw[white, fill=white] (-0.0625,0.095) 
	circle [radius=0.007];
	\end{tikzpicture}
	\hspace{-2mm}
}

\foreach \x in {A, ..., Z}{
	\expandafter\xdef\csname orcid\x\endcsname{\noexpand\href{https://orcid.org/\csname orcidauthor\x\endcsname}{\noexpand\orcidicon}}
}

\usepackage[inline]{./trackchanges}

\begin{document}

\title{User-Movement-Robust Virtual Reality Through Dual-Beam Reception in mmWave Networks}

\author[1]{Rizqi Hersyandika}
\author[2]{Qing Wang}
\author[1,3]{Yang Miao}
\author[1]{Sofie Pollin}

\affil[1]{KU Leuven, Belgium}
\affil[2]{Delft University of Technology, The Netherlands}
\affil[3]{University of Twente, The Netherlands}

\maketitle

\begin{abstract}

Utilizing the mmWave band can potentially achieve the high data rate needed for realistic and seamless interaction within a virtual reality (VR) application. To this end, beamforming in both the access point (AP) and head-mounted display (HMD) sides is necessary. The main challenge in this use case is the specific and highly dynamic user movement, which causes beam misalignment, degrading the received signal level and potentially leading to outages. This study examines mmWave-based coordinated multi-point networks for VR applications, where two or multiple APs cooperatively transmit the signals to an HMD for connectivity diversity. Instead of using omni-reception, we propose dual-beam reception based on the analog beamforming at the HMD, enhancing the receive beamforming gain towards serving APs while achieving diversity. Evaluation using actual HMD movement data demonstrates 
the effectiveness of our approach, showcasing a reduction in outage rates of up to 13\% compared to quasi-omnidirectional reception with two serving APs, and a 17\% decrease compared to steerable single-beam reception with a serving AP. Widening the separation angle between two APs can further reduce outage rates due to head rotation as rotations can still be tracked using the steerable multi-beam, albeit at the expense of received signal levels reduction during the non-outage period.

\end{abstract}


\IEEEpeerreviewmaketitle

\section{Introduction}
\label{sec_introduction}

\subsection{Motivations}

Immersive virtual reality (VR) applications hold huge promises for industries such as automotive, healthcare, education and tourism. Achieving truly immersive experiences relies heavily on high-speed data transfer, which is essential for rendering realistic environments, facilitating smooth interactions, and enabling seamless content streaming. Such high data rates can be easily achieved when operating in the millimeter-wave (mmWave) frequency bands, thanks to the abundant bandwidth provided in these bands~\cite{Struye2020}. Nevertheless, the communication in these bands suffers from high path loss compared to the lower frequency bands, requiring the beamforming gain compensation in both transmission and reception.

Addressing user movement is crucial when operating the VR in mmWave networks~\cite{Struye2022}, as user movement can cause beam misalignment leading to channel gain reduction. An inertial measurement unit (IMU) sensor-assisted beam realignment for the mmWave user device has been proposed in~\cite{Bao2018, Lin2019} as a user-side solution to compensate for the beam misalignment caused by the user movement, and to perform beam-tracking. This approach is also applicable to the VR head-mounted display (HMD), which is commonly equipped with the IMU sensor. Another approach employs an interleaved sub-array on the HMD, generating a sub-beam to proactively predict the future angle-of-arrival due to movement~\cite{Struye2023}. However, this approach suffers from the undesired mutual coupling effect, degrading the array performance. Experimental investigations on the impact of array configuration on HMDs have been conducted by~\cite{Alex2023}.  

Several solutions on the mmWave access point (AP) incorporate the mmWave network with 1) the legacy sub-6~GHz Wi-Fi as a fall-back mechanism~\cite{Gupta2019,Peng2022} and 2) the free-space optics for dual-connectivity~\cite{Blandino2023} to mitigate mmWave link outage due to user movement. The use of mmWave coordinated multi-point (CoMP) networks investigated in~\cite{Peng2022} enhances reliability by providing spatial diversity to serve VR users. These networks mitigate single-link outages due to user movement by transmitting the same stream through multiple spatially separated APs. Multiple transmit (Tx) beams can be generated either through spatially separated coordinated APs~\cite{Maamari2016} or by a single AP transmitting two synchronized beams: a line-of-sight (LoS) beam and a reflected one, providing two constructive beams at the user device if the path delays are sufficiently compensated~\cite{Jain2021}. However, these earlier studies only consider omnidirectional or quasi-omni reception at the receive (Rx) side, resulting in low channel gain. This motivates us to integrate the mmWave CoMP networks with multi-beam reception at the HMD side to improve the communication channel gain. 

Multi-beam reception can be straightforwardly achieved by employing hybrid beamforming at the HMD, where multiple radio frequency (RF) chains control subsets of antenna elements to generate independent Rx beams directed at different APs. Nevertheless, employing multiple RF chains at the HMD may not be efficient, as it would receive only one data stream from multiple Tx directions. 
Moreover, employing multiple RF chains on the mmWave user side demands substantial energy consumption for both operation and processing of high-bandwidth signals~\cite{Alkhateeb2014_mag}, especially considering the limited battery capacity. Consequently, prior studies have predominantly considered analog beamforming at the mmWave user device.

\subsection{Contributions}

The contributions of this paper are listed as follows:
\begin{itemize}
    \item We introduce a dual-beam reception approach for VR applications, realized through analog beamforming using a uniform planar array (UPA) at the HMD. This approach enables highly directional receptions from multiple coordinated serving APs, enhancing the channel gain and improving the robustness against rapid head movement during VR streaming. 

    \item We derive a beam misalignment model between the HMD and serving AP(s), converting the 6 degrees of freedom (6DOF) head movement into position displacement and beam misalignment in azimuth and elevation directions. Subsequently, we emulate the channel between serving APs and a freely moving HMD using real HMD movement data from~\cite{Lo2017}. 

    \item We assess the performance of dual-beam reception across various separation angles between two serving APs. Our dual-beam reception approach outperforms the fixed quasi-omni reception and the steerable single-beam reception in terms of channel gain. In addition, our findings indicate that increasing the separation angle between APs can reduce the outage rate caused by head rotation. However, this improvement comes at the cost of reduced achievable received signal level during non-outage periods, attributed to the reduced beamforming gain of multi-beams steered away from the boresight.
\end{itemize}

\section {System Model}
\label{sec:system_model}

\begin{figure}[t]
\centering
\includegraphics[width=0.64\linewidth]{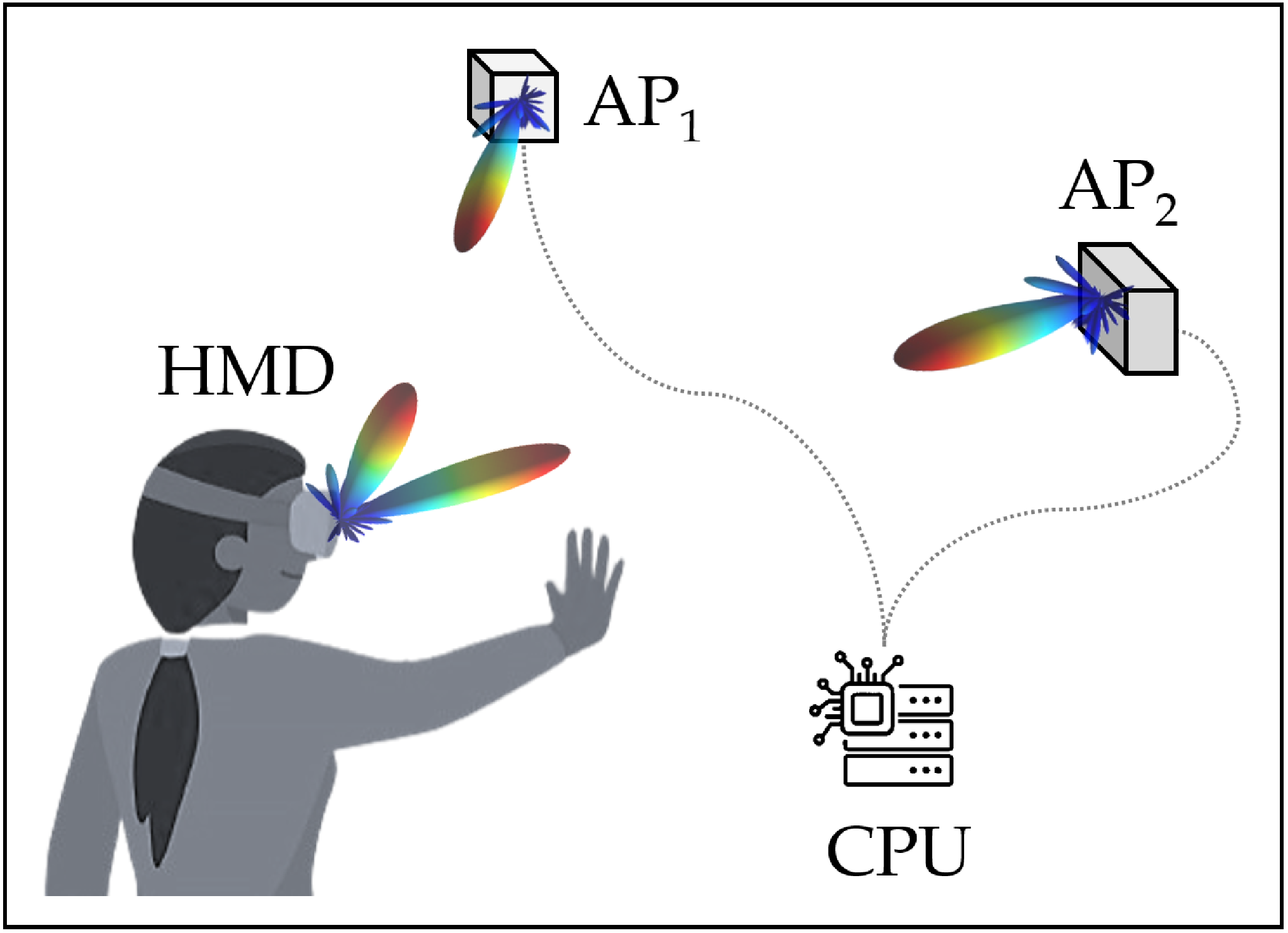}
\caption{\centering Dual-beam reception at HMD-VR in mmWave \ CoMP networks}
\label{fig:HMD_VR}
\end{figure}

We consider a mmWave coordinated multi-point network. As illustrated in Fig.~\ref{fig:HMD_VR}, multiple spatially distributed APs, coordinated by a central processing unit (CPU), cooperate to serve an HMD user using multiple directive Tx beams. The HMD employs multiple directive Rx beams aligned with the Tx beam directions. It is crucial for the HMD employing the analog multi-beam reception to receive signals from $L$ serving APs that add constructively. Hence, we consider maximum ratio transmission (MRT) precoding to transmit downlink signals from $L$ serving APs, ensuring their constructive reception at the HMD. A more advanced precoding technique such as zero-forcing (ZF) or minimum mean square error (MMSE) could be used to mitigate the interference in the presence of multiple HMD users.

During the channel estimation phase, the HMD transmits an uplink pilot to all $L$ serving APs within its field of view. The received uplink pilot at AP-$l$ can be expressed as: 
\begin{equation}\label{eq:y_ul}
    y_{l}^{\text{UL}} = \sqrt{p_{k}} \hspace{0.05cm} \hat{h}_{l}^{\text{UL}} \hspace{0.05cm} {s}_{p} + \hat{z}_{l},
\end{equation}
where $p_k$ denotes the transmit power of HMD, $\hat{h}_{l}^{\text{UL}}$ represents the uplink channel between the HMD and AP-$l$, and ${s}_p$ being the transmitted pilot symbol. $\hat{z}_{l}$ represents the additive complex white Gaussian noise at AP-$l$ with zero mean and variance $\sigma_l^2 = k_B \hspace{0.08cm} T \hspace{0.02cm} B \hspace{0.02cm} F_{l}$, with $k_B$ being the Boltzmann constant, $T$ denoting the temperature, $B$ representing the system bandwidth and $F_{l}$ being the noise figure of the AP-$l$. 
We assume perfect channel estimation between HMD and AP-$l$, thus $\hat{h}_{l} = \hat{h}_{l}^{\text{UL}}$. The CPU then collects and combines all estimated channels between the serving APs and the HMD, denoted as $\hat{H} = [\hat{h}_1, ... , \hat{h}_L]$. Assuming channel reciprocity in Time Division Duplex (TDD), $\hat{h}_{l}^{\text{UL}} = \hat{h}_{l}^{\text{DL} *}$, the normalized MRT precoding vector for AP-$l$ can be expressed as:
\begin{equation}
    \hat{v}_l = \frac{\hat{h}_l}{|\hat{h}_l|}.
\end{equation}
The overall precoding vector assigned by the CPU is denoted as $\hat{V} = [\hat{v}_1, ... , \hat{v}_L]$.

On the downlink side, we consider the dominant LoS paths between the HMD and serving APs. The received signals at the HMD are the coherent summation of all downlink signals transmitted from $L$ serving APs, formulated as:
\begin{equation}
\label{eqn:sum_dl_transmission}
    \hat{y}^{\text{DL}} = \sum\limits_{l = 1}^{L} \sqrt{p_{l}} \hspace{0.05cm} \hat{v}_{l} \hspace{0.05cm} \hat{h}_{l}^{\text{DL}} \hspace{0.05cm} {s}_{d} \hspace{0.05cm} + \hat{z}_{k}, 
\end{equation}
where $p_{l}$ denotes the transmit power of AP-$l$, $\hat{v}_{l}$ is the precoding vector of AP-$l$, $\hat{h}_{l}^{\text{DL}}$ represents the complex downlink channel between AP-$l$ and the HMD, ${s}_{d}$ is the transmitted data stream, and $\hat{z}_{k}$ denotes the additive white Gaussian noise at the HMD-$k$ with the noise figure of $F_{k}$.

The complex LoS downlink channel between AP-$l$ and the HMD includes the beamforming gain at both the AP and HMD, and can be expressed as: 
\begin{equation}\label{eq:h_ul}
    \hat{h}_{l}^{\text{DL}} = \frac{c}{4\pi f_{c} \hspace{0.05cm} d_{l}} \hspace{0.05cm} \sqrt{ g_{l}^{\text{tx}}(\theta_{D},\phi_{D}) \hspace{0.1cm} g^{\text{rx}}(\theta_{A},\phi_{A}) } \hspace{0.05cm} e^{-j \frac{2 \pi f_{c} d_{l}}{c} }, 
\end{equation}
where $f_{c}$ denotes the carrier frequency, $c$ represents the speed of light, $d_{l}$ is the distance between the HMD and AP-$l$, $g_{l}^{\text{tx}}(\theta_D,\phi_D)$ denotes the Tx beamforming gain of AP-$l$ at the angle of departure (AoD) of $(\theta_D,\phi_D)$, and $g^{\text{rx}}(\theta_A,\phi_A)$ denotes the Rx beamforming gain of HMD at the angle of arrival (AoA) of $(\theta_A,\phi_A)$.

\section {Beam Misalignment Model}
\label{sec:misalignment_model}

\subsection {Translational motion}

The translational movement of an HMD not only changes the HMD--AP distance but also causes beam misalignment between them, as the positional displacement of the HMD changes its orientation with respect to the serving AP. 
We denote the position of HMD-$k$ and AP-$l$ in Cartesian coordinates as ${P}_{k_{x,y,z}} = [p_{k_x},p_{k_y},p_{k_z}]$ and ${P}_{l_{x,y,z}} = [p_{l_x},p_{l_y},p_{l_z}]$, respectively. 
The orientation vector between HMD-$k$ and AP-$l$ can be expressed as: 
\begin{equation}
    Q_{r,\theta,\phi} = [r_{kl}, \theta_{kl}, \phi_{kl}]. 
\end{equation}
$r_{kl}$ denotes the HMD--AP distance, while $\theta_{kl}$ and $\phi_{kl}$ represent the HMD's relative azimuth and elevation angles with respect to the AP, formulated as:
\begin{equation}
    r_{kl} = \sqrt{{(p_{k_x} - p_{l_x})}^2 + {(p_{k_y} - p_{l_y})}^2 + {(p_{k_z} - p_{l_z})}^2},   
    \label{eq:dist}
\end{equation}
\begin{equation}
    \theta_{kl} = \arctan \left( \frac {p_{k_y} - p_{l_y}} {p_{k_x} - p_{l_x}} \right), 
    \label{eq:az}
\end{equation}
\begin{equation}
    \phi_{kl} = \arctan \left( \frac {\left(p_{k_z} - p_{l_z}\right)} {\sqrt{{(p_{k_x} - p_{l_x})}^2 + {(p_{k_y} - p_{l_y})}^2}} \right). 
    \label{eq:el}
\end{equation}

Denoting the HMD's translational movement in the corresponding axis as $p'$, as illustrated in Fig.~\ref{fig:head_mvmt}, the HMD-$k$'s position after the translational movement can be expressed as: 
\begin{equation}
    {P}_{k_{x,y,z}}^{\text{trn}} = \begin{bmatrix} p_{k_x}^{\text{trn}} \\ p_{k_y}^{\text{trn}} \\ p_{k_z}^{\text{trn}} \end{bmatrix} = \begin{bmatrix} p_{k_x} + p_{k_x}'\\ p_{k_y} + p_{k_y}' \\ p_{k_z} + p_{k_z}' \end{bmatrix}.
\end{equation}
By substituting ${P}_{k_{x,y,z}}^{\text{trn}}$ to (\ref{eq:dist}), (\ref{eq:az}) and (\ref{eq:el}), the resulting HMD orientation vector after translational movement is written as:
\begin{equation}
    Q_{r,\theta,\phi}^\text{trn} = [r_{kl}^{\text{trn}}, \theta_{kl}^{\text{trn}}, \phi_{kl}^{\text{trn}}].
    \label{eq:q_trans}
\end{equation}

\subsection {Rotational motion}

\begin{figure}[t]
\centering
\includegraphics[width=0.6\linewidth]{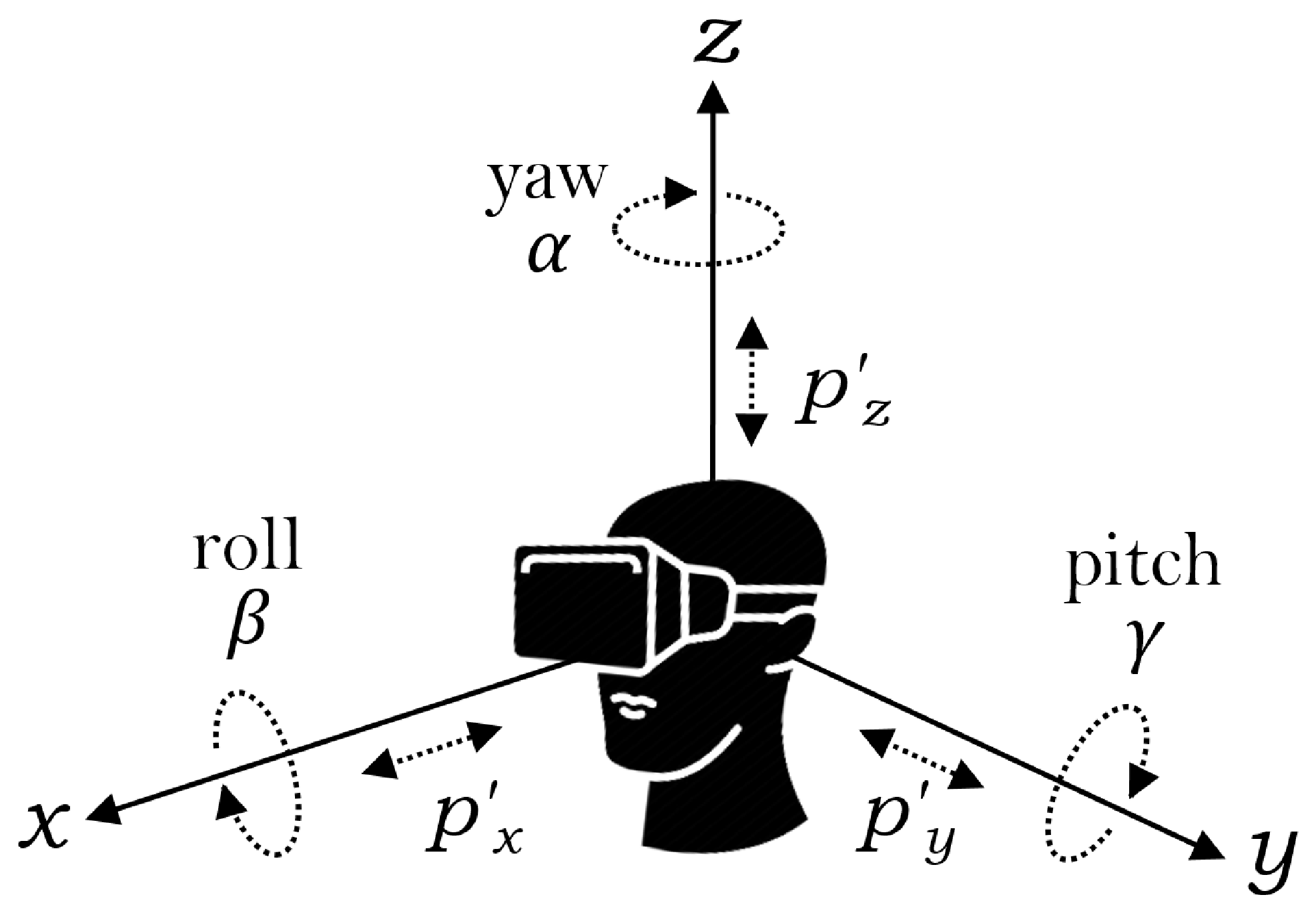}
\caption{The 6DOF head movement illustration} 
\label{fig:head_mvmt}
\end{figure}

The HMD rotation only changes the HMD's orientation with respect to the AP without changing the HMD--AP distance, assuming the distance between the head center to VR antennas is negligible compared to the head-to-AP distance. In the IMU sensor, the device rotation is typically represented by yaw, pitch, and roll movements, as depicted in Fig.~\ref{fig:head_mvmt}. Yaw signifies left and right head rotations around the vertical $z$-axis (i.e., shaking head), pitch indicates up and down head movements around the $y$-axis (i.e., nodding), and roll expresses head-tilt movements around the $x$-axis. These rotational movements can be translated into the following Euler rotation matrix:
\begin{equation}
    {R}_x = \begin{bmatrix}
            1 & 0 & 0 \\
            0 & \cos(\gamma) & \sin(\gamma) \\
            0 & -\sin(\gamma) & \cos(\gamma)
        \end{bmatrix}
\end{equation}
\begin{equation}
    {R}_y = \begin{bmatrix}
            \cos(\beta) & 0 & -\sin(\beta) \\
            0 & 1 & 0 \\
            \sin(\beta) & 0 & \cos(\beta)
        \end{bmatrix}
\end{equation}
\begin{equation}
    {R}_z = \begin{bmatrix}
            \cos(\alpha) & \sin(\alpha) & 0 \\
            -\sin(\alpha) & \cos(\alpha) & 0 \\
            0 & 0 & 1
        \end{bmatrix},
\end{equation}
where $\alpha$, $\beta$, and $\gamma$ correspond to yaw, pitch, and roll angle movements, respectively. Following the $XYZ$ convention rotation, the rotation matrix becomes: 
\begin{equation}
    {R} = {R}_x {R}_y {R}_z.
\end{equation}    
Thus, the rotated vector in the $x,y,z$ plane becomes:
\begin{equation}
    {Q}_{x,y,z}^{\text{rot}} = {Q}_{x,y,z} {R}, 
\end{equation}
where ${Q}_{x,y,z}^{\text{rot}} = [{q_{x}^{\text{rot}}}, {q_{y}^{\text{rot}}}, {q_{z}^{\text{rot}}}]$. The orientation vector after HMD rotation is converted to the spherical format following:  
\begin{equation}
    {Q}_{r,\theta,\phi}^{\text{rot}} = \begin{bmatrix} \sqrt{ {q_{x}^{\text{rot}}}^2 + {q_{y}^{\text{rot}}}^2 + {q_{z}^{\text{rot}}}^2} \\ \arctan \left( {q_{y}^{\text{rot}}} \left({q_{x}^{\text{rot}}} \right)^{-1} \right) \\  \arctan {\left( {q_{z}^{\text{rot}}} { \left(\sqrt{ {q_{x}^{\text{rot}}}^2 + {q_{y}^{\text{rot}}}^2} \right)}^{-1} \right)}  \end{bmatrix}.    
\end{equation} 

The misalignment orientation vector due to combined translation and rotation movements with respect to the initial orientation vector ${Q}_{r,\theta,\phi}$ is formulated as:
\begin{equation} 
    {Q}_{r,\theta,\phi}^{\text{mis}} = 2 {Q}_{r,\theta,\phi} - \left ( {Q}_{r,\theta,\phi}^{\text{trn}} + {Q}_{r,\theta,\phi}^{\text{rot}} \right).
\end{equation}
${Q}_{r,\theta,\phi}^{\text{mis}} = [r^{\text{dis}}, \theta^{\text{mis}}, \phi^{\text{mis}}]$, where $r^{\text{dis}}$ represents the HMD--AP distance discrepancy due to displacement, $\theta^{\text{mis}}$ and $\phi^{\text{mis}}$ denote the misalignment in the azimuth and elevation directions, respectively.

\section{Dual-beam Reception Beamforming}
\label{sec_beamforming}

\subsection {Beamforming with UPA}
We consider a UPA$_{(M \times N)}$ with the size of $M \times N$ antenna elements employed at each AP and HMD side. The antenna steering vector of the $m$-th horizontal element and $n$-th vertical element for the azimuth angle $\theta$ and the elevation angle $\phi$ are expressed as: 
\begin{equation}
    \hat{a}_{m}(\theta,\phi) = e^{-j \frac{2 \pi}{\lambda} \hspace{0.02cm} m \hspace{0.02cm} d_x \hspace{0.02cm} \sin \theta \hspace{0.02cm} \cos \phi},
\end{equation}
\begin{equation}
    \hat{a}_{n}(\theta,\phi) = e^{-j \frac{2 \pi}{\lambda} \hspace{0.02cm} n  \hspace{0.02cm} d_y \hspace{0.02cm} \sin \theta \hspace{0.02cm} \sin \phi}, 
\end{equation}
respectively. The steering vector is assumed to be a function of a single frequency. The steering vector of the ($m, n$)-th elements of UPA is~\cite{Tan2017}:
\begin{equation}
    \hat{a}_{m,n}(\theta,\phi) = \hat{a}_{m}(\theta,\phi) \hat{a}_{n}(\theta,\phi), 
\end{equation}
where $m = \{0, ..., M-1\}$ and $n = \{0, ..., N-1\}$. By considering a half-wavelength spacing between two horizontal elements $d_x$ and vertical elements $d_y$, the steering vector becomes:
\begin{equation}
    \hat{a}_{m,n}(\theta,\phi) = e^{-j \pi \sin \theta (m \cos \phi + n \sin \phi)}.
\end{equation}
The overall array steering vector is denoted as $\hat{\mathcal{A}} \in \mathbb{C}^{1 \times MN}$, where $ \hat{\mathcal{A}} = \{\hat{a}_{1,1},...,\hat{a}_{M,N} \}$. 

To steer the beam toward a specific direction, a beamforming weight vector needs to be assigned to each element. The weight vector of the ($m, n$)-th element for steering the beam in the direction $(\theta_b, \phi_b)$ can be defined as follows: 
\begin{equation}
    \hat{w}_{m,n}(\theta_b,\phi_b) = e^{j \pi \sin \theta_b (m \cos \phi_b + n \sin \phi_b)}.
\end{equation}
The overall beamforming weight vector of an array is denoted as $\hat{\mathcal{W}} \in \mathbb{C}^{1 \times MN}$, where $ \hat{\mathcal{W}} = \{\hat{w}_{1,1},...,\hat{w}_{M,N} \}$. 
The beamforming gain in the direction $(\theta, \phi)$ is expressed as:
\begin{equation}
\label{eq:gain_bf}
    g(\theta, \phi) = \left| \hat{\mathcal{W}}^\text{T}(\theta_b,\phi_b) \hat{\mathcal{A}}(\theta,\phi) \right|^2.
\end{equation}

\subsection{Dual-beam reception at the HMD}

\begin{figure}[t]
\centering
\includegraphics[width=0.84\linewidth]{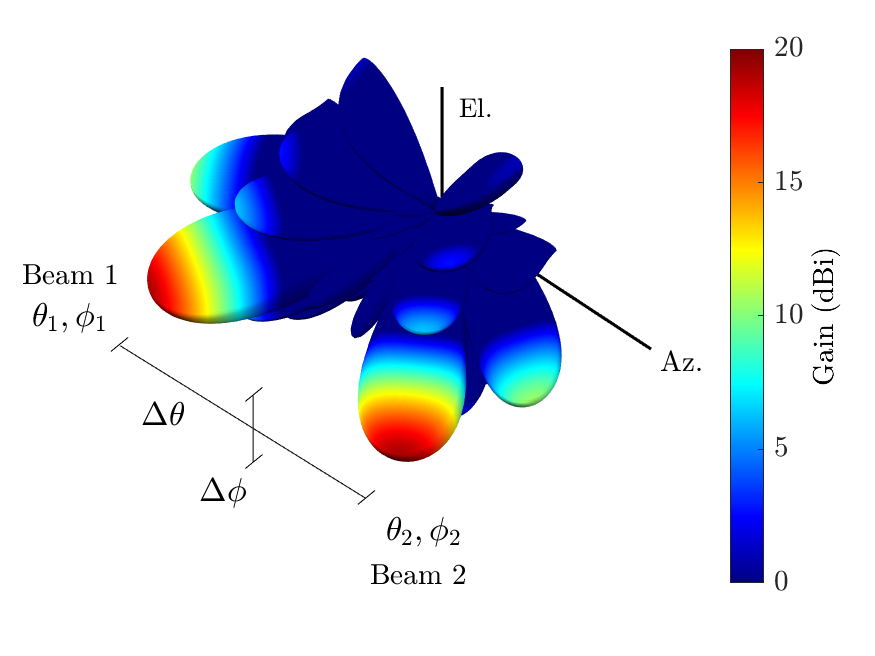}
\caption{Multi-beam pattern for HMD reception}
\label{fig:multibeam_pattern}
\end{figure}

The HMD applies dual-beam reception to receive the signals from multiple serving APs with strong gains, steering two directional Rx beams towards the APs' directions. These multi-directional beams are realized by combining multiple beamforming weight vectors in the directions of $L$ serving APs. In general, the combined beamforming weight vectors assigned by the beamformer to generate $L$ directive beams can be expressed as:
\begin{equation}
    \label{eq:w_multibeam}
    \hat{\mathcal{W}}_{L} = \frac{ \sum_{l=1}^{L} \sqrt{\eta_{l}} \hat{\mathcal{W}}_{l}(\theta_l,\phi_l)}{\sum_{l=1}^{L} \left|\sqrt{\eta_{l}} \hat{\mathcal{W}}_{l}(\theta_l,\phi_l)\right|},
\end{equation}
where $\hat{\mathcal{W}}_{l}(\theta_l,\phi_l)$ denotes the beamforming weight vector of UPA to beamform towards the intended $(\theta_l,\phi_l)$ direction, and $\eta_l$ represents the power coefficient of beam-$l$, where $\sum_{l=1}^{L}\eta_l=1$.  
The beamforming gain of multi-beam is then obtained by substituting (\ref{eq:w_multibeam}) to (\ref{eq:gain_bf}). 

Fig.~\ref{fig:multibeam_pattern} illustrates the beam pattern of dual-beam ($L = 2$), where the first beam is steered at ($\theta_1,\phi_1$) and the second beam is steered at ($\theta_2,\phi_2$), generated using UPA$_{(8\times8)}$. The separation in the azimuth and elevation directions are denoted as $\Delta\theta = |\theta_2 - \theta_1|$ and $\Delta\phi = |\phi_2 - \phi_1|$. In this example, the power among the two beams is shared equally with $\eta_1 = \eta_2 = 0.5$, resulting in identical beamforming gain in both intended steering directions, $g(\theta_1,\phi_1) = g(\theta_2,\phi_2)$.

\section{Results and Analysis}
\label{sec:results}

\subsection{Simulation setup and HMD movement emulation}
\label{subsec:simulation_setup}

\begin{figure}[t]
\centering
\includegraphics[width=0.47\linewidth]{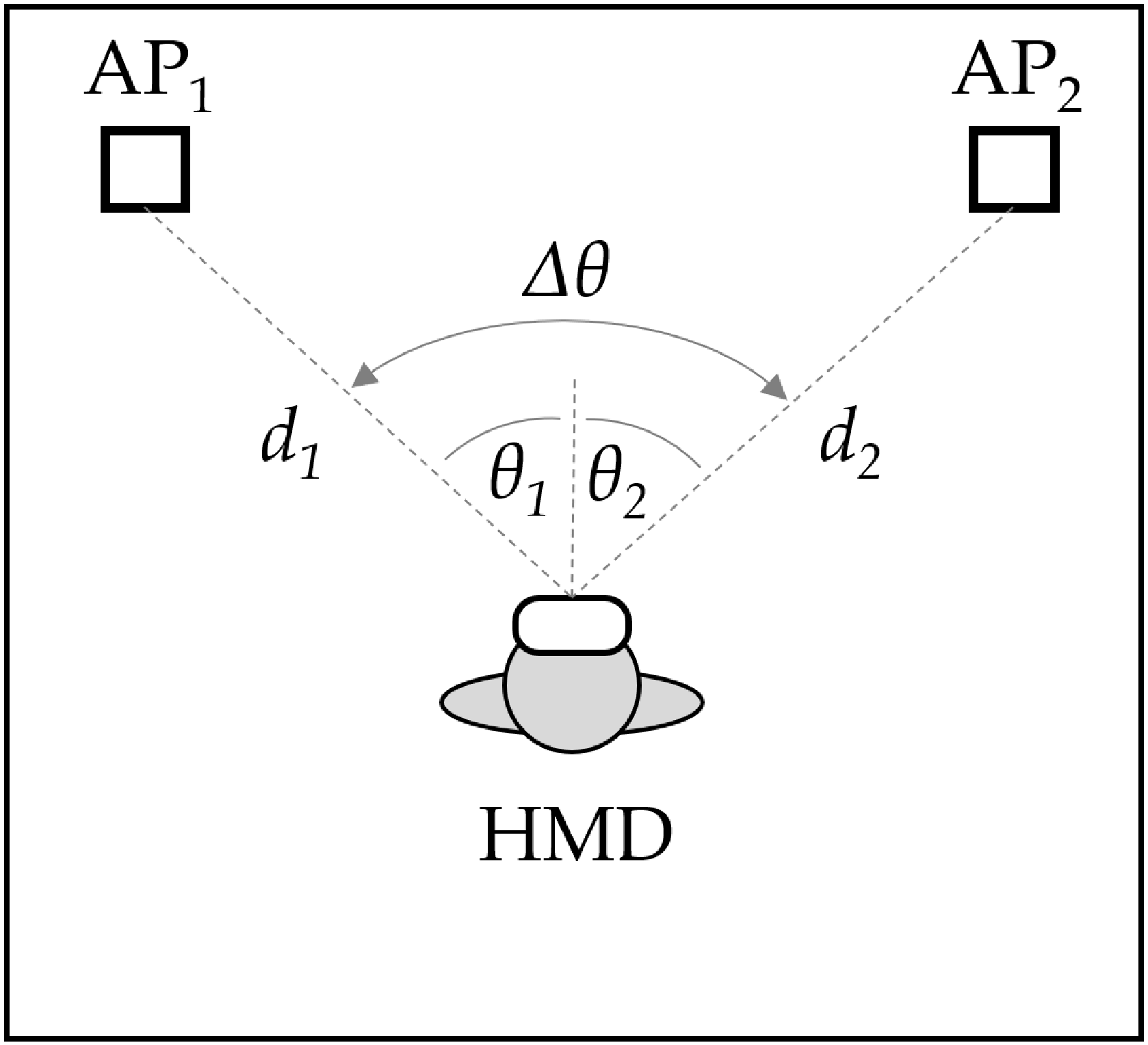}
\caption{\centering Simulation setup}
\label{fig:sim_setup}
\end{figure}

We simulate the scenario of an indoor VR room with a size of $20\times20$~m$^2$. To evaluate the performance of HMD dual-beam reception during the user movement, we deploy two APs at a height of 4~m symmetrically on one side of the room. For the initial observation point, an HMD with a height of 1.5~m is positioned around the center of a room, resulting in two identical distances ($d_1 = d_2$) and azimuth angles ($\theta_1 = \theta_2$) between the HMD and the two APs, as illustrated in Fig.~\ref{fig:sim_setup}. The separation angle between two APs is denoted as $\Delta\theta$. We maintain the same HMD--AP distances $d_1 = d_2 = 10$~m at the initial alignment and vary the $\Delta\theta$ values to analyze the impact of the AP separation on signal reception performance.
The placement of serving APs and the VR user is based on the typical scenario where during initial VR setup, the user stands at 
an ideal position within the coverage of serving APs, with the HMD predominantly facing one side during streaming. 

To emulate the HMD movement over time, we use the 360$\degree$ HMD movement dataset provided by~\cite{Lo2017}. This dataset contains the 6DOF movement of an HMD, sampled every $\sim$40~ms and collected from 50 participants watching 10 video themes. For our simulation, we selected all video themes collected from 3 participants, each with a duration of 60~s. We sample the data every 320~ms and assume that the beam alignment process between APs and HMD is accomplished within each time step. We then transform the HMD movement into beam misalignment and simulate the channel between APs and HMD over time. 

Each AP transmits signals using a carrier frequency of 28~GHz with a bandwidth of 200~MHz. The transmit power is set at 10~dBm, while HMD and AP have a noise figure of 7~dB. We employ UPA$_{(8\times8)}$ at both HMD and AP sides, allowing both to steer their beams according to the HMD's position and orientation. Each AP generates a single directive beam, while the HMD generates dual-beam reception using (\ref{eq:w_multibeam}). As a baseline, we consider a quasi-omni reception at the HMD, realized using a smaller array size, UPA$_{(2\times4)}$. It generates a wider Rx beam in the azimuth and elevation directions, with a fixed beam steered at the boresight, $(\theta_{b},\phi_{b}) = (0\degree,0\degree)$.

\subsection{Dual-beam gain analysis}
\label{subsec:gain_analysis}

We evaluate the beamforming gain of the two steered beams over azimuth angle separation $\Delta\theta$. The azimuth angle of the first beam is denoted as $\theta_{1}$ and the azimuth angle of the second beam is $\theta_{2} = \theta_{1} - \Delta \theta$. The elevation angle separation is set to $\Delta \phi = 0\degree$, where $\phi_{1} = \phi_{2} = 0\degree$.  We consider the equal power distribution among the two beams ($\eta_1 = \eta_2 = 0.5$), resulting in identical beamforming gains in both intended steering directions, $g(\theta_1,\phi_1) = g(\theta_2,\phi_2)$. 

\begin{figure}[t]
\centering
\includegraphics[width=0.71\linewidth]{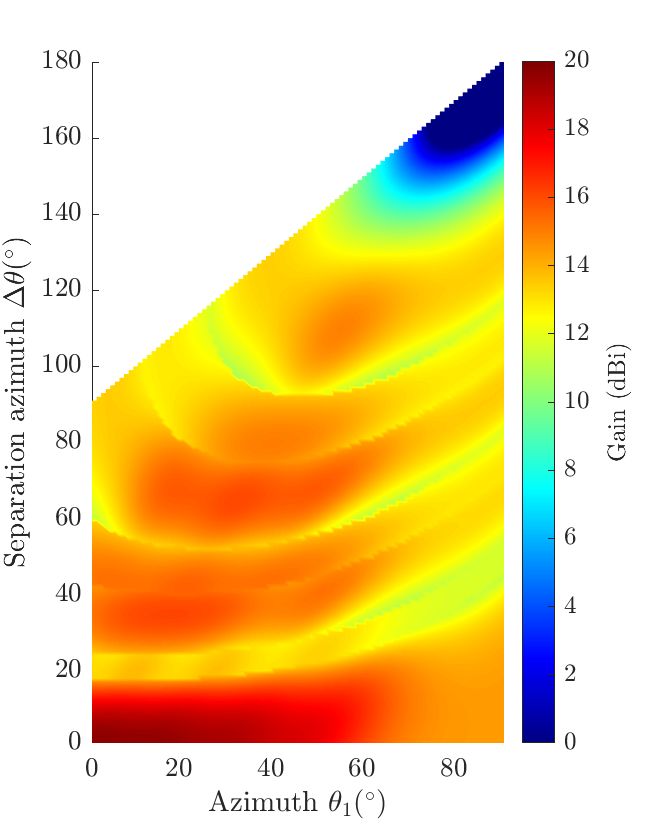}
\caption{\centering The beamforming gain of dual-beam over separation azimuth angles with $\phi_{1} = \phi_{2} = 0\degree$ and $\eta = 0.5$}
\label{fig:sep_angle_gain}
\end{figure}

Fig.~\ref{fig:sep_angle_gain} presents the beamforming gain of the dual-beam over $\theta_{1}$ and $\Delta\theta$. It shows that the maximum gain is achieved at $\theta_{1} = 0\degree$ and $\Delta\theta = 0\degree$, indicating when there is only one single beam steered at the azimuth angle of $0\degree$. Splitting into dual-beam reduces the achieved beamforming gain in each intended direction. Nevertheless, the gain reduction shows a non-linear relation with the increasing $\Delta\theta$ in a fixed $\theta_{1}$, as certain gain patterns are observed in Fig.~\ref{fig:sep_angle_gain}. Significant gain reduction is observed when both beams are steered away from the boresight and are separated with a large $\Delta\theta$.

\subsection{Performance during the HMD movement}
\label{subsec:performance}

\begin{figure}[t]
\centering
\includegraphics[width=0.99\linewidth]{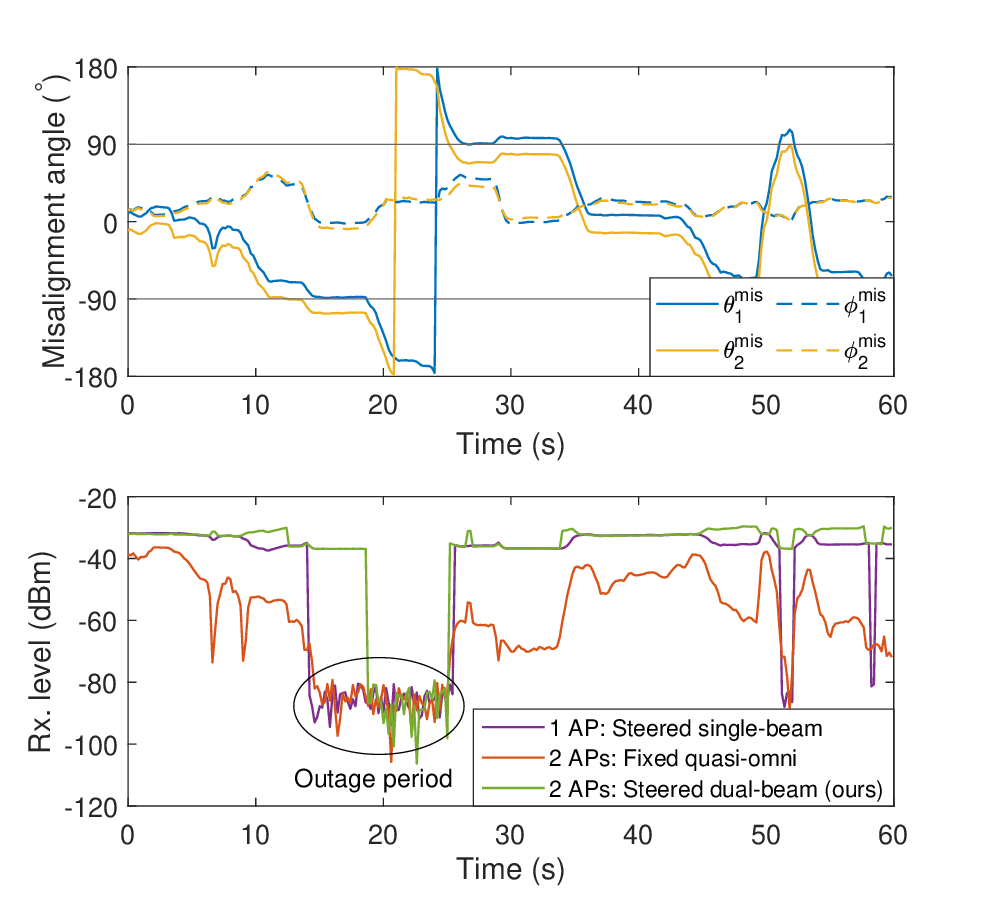}
\caption{\centering The impact of misalignment on the Rx. level}
\label{fig:mis_rxlevel}
\end{figure}

Fig.~\ref{fig:mis_rxlevel} shows the impact of HMD movement on the Rx signal level, evaluated from one of the datasets. The top subfigure illustrates the misalignment angle in the azimuth and elevation plane, denoted as $\theta^{\text{mis}}$ and $\phi^{\text{mis}}$, with respect to the initial orientation between HMD and serving APs, evaluated over a 60~s period. The most significant misalignment occurs in the azimuth directions, primarily due to yaw head rotation. The discrepancy between $\theta_1^{\text{mis}}$ and $\theta_2^{\text{mis}}$ at each time step is due to the separation angle between the two APs. 

The bottom subfigure of Fig.~\ref{fig:mis_rxlevel} describes the Rx level resulting from HMD movement. Reception using a fixed quasi-omni pattern yields a significantly lower and more fluctuating Rx level compared to the steered beam, even when two APs are employed to serve the HMD. With periodic beam re-alignment in the case of steered single-beam and dual-beam, the Rx level can be maintained as long as the serving AP remains in the field of view of the HMD. Outages occur when the HMD rotates beyond $\pm90\degree$ from an AP, causing no AP capable of serving it given that mmWave diffraction over the head does not contribute to the Rx signal. The Rx level recovers once it's back within the field of view of the HMD. Relying on only one AP leads to outages occurring shortly after the AP is out of the field of view of the HMD. By distributing more APs, the outage period can be reduced as one of the APs will remain serving the HMD for an extended duration.

\begin{figure}[t]
\centering
\includegraphics[width=0.98\linewidth]{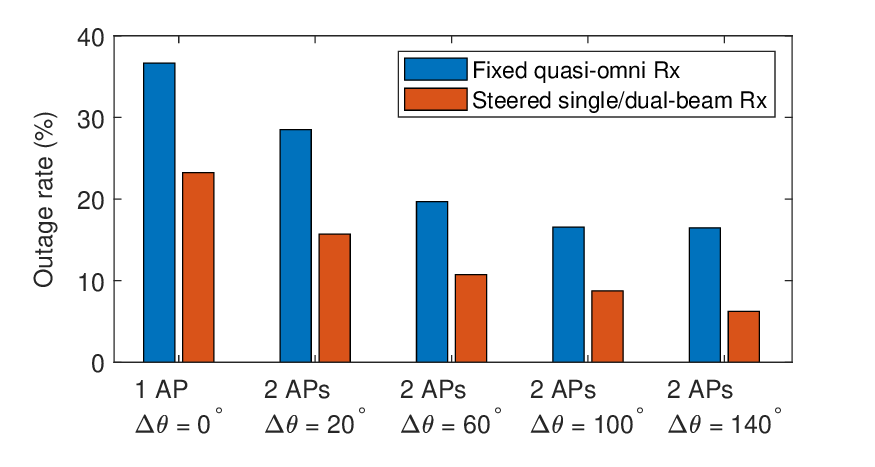}
\caption{Outage rate due to HMD movement}
\label{fig:outage_rate}
\end{figure}

We evaluate the impact of AP separation on the outage rate. The outage rate metric, defined as the total outage duration over total length of the observation period, is evaluated over different azimuth separations $\Delta\theta$ between APs, as presented in Fig.~\ref{fig:outage_rate}. For all cases, using steerable beam reception reduces the outage rate by up to 13\% compared to fixed quasi-omni reception. The outage rate also decreases as $\Delta\theta$ increases, with up to 17\% lower outage period compared to single-beam reception when $\Delta\theta = 140\degree$. A larger separation between APs allows the HMD to be covered by at least one AP most of the time during rotation, thereby reducing the outage rate.   

\begin{figure}[t]
\centering
\includegraphics[width=0.98\linewidth]{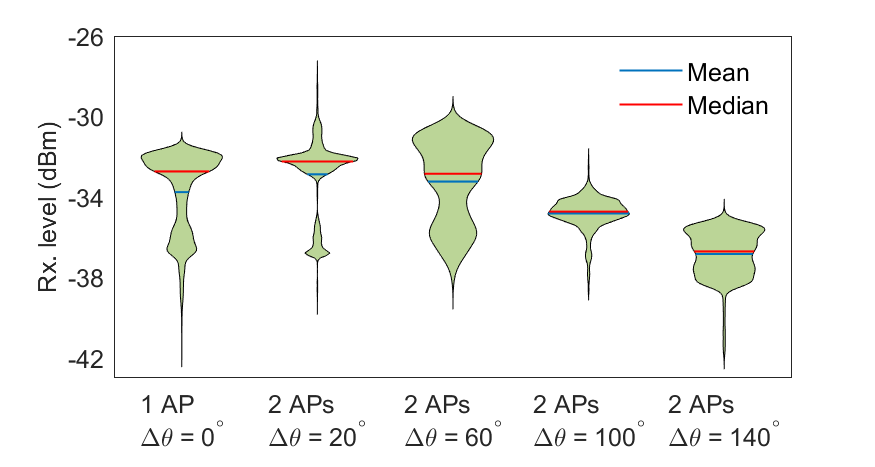}
\caption{Rx. level during the aligned (non-outage) period}
\label{fig:rxlevel_nonoutage}
\end{figure}

We also assess the Rx signal level at the HMD during the aligned (non-outage) period. The violin plots in Fig.~\ref{fig:rxlevel_nonoutage} show the distribution of the Rx signal level for various AP separation angles. An average Rx level improvement over the single-beam case is observed at azimuth separations of $20\degree$ and $60\degree$, demonstrating the advantage of spatial diversity and the dual-beam reception gain. The highest achievable Rx level at $\Delta\theta = 20\degree$ results from the two Rx beams steered in the direction around the boresight angle of the HMD. At wider separation angles (i.e., $\Delta\theta = 100\degree$ and $140\degree$), the overall Rx level performance is not superior to the single-beam case. This is primarily attributed to the reduction in both transmit and receive beamforming gains, as beams steered away from the boresight exhibit wider beamwidths. Therefore, there exists a trade-off between the robustness of widely separated APs in minimizing outage rates and the Rx signal level experienced during the aligned period. These insights are crucial for planning AP deployment and/or designing AP selection algorithms for mmWave CoMP networks serving the HMD with dual or multi-beam reception capability.

\section{Conclusion}
\label{sec_conclusion}

This paper proposes a user-movement-robust VR approach by integrating mmWave coordinated multi-point networks and steerable dual-beam reception at the HMD. The proposed dual-beam reception HMD, realized by combining multi-directional analog beamforming weight vectors using UPA, significantly enhances the channel gain compared to omni-reception HMDs, as they allow exploiting an array gain. The analysis of the dual-beam gain indicates a non-linear relationship between the azimuthal separation of beams and the achieved beamforming gain. Through emulating the channel based on real HMD movement data, we observe the received signal outage when the serving AP is out of view due to head rotation. Employing two widely separated serving APs with steerable dual-beam reception at the HMD reduces the outage rate at the cost of reducing the received signal level during the aligned period. This underscores the importance of strategically distributing serving APs around the user to prevent outages from rapid user rotation. 
Our future work will consider using more distributed APs in multi-HMD scenarios,  including optimizing beam power distribution in multi-directional analog beamforming.

\section*{Acknowledgment}
This work has received funding from the Smart Networks and Services Joint Undertaking (SNS JU) under the European Union’s Horizon Europe research and innovation programme under Grant Agreement No. 101096954 (6G-BRICKS project) and No. 101139291 (iSEE-6G project).

\bibliographystyle{IEEEtran}
\bibliography{./mybib}

\end{document}